\def\be{\begin{equation}}
\def\ee{\end{equation}}
\def\bea{\begin{eqnarray}}
\def\eea{\end{eqnarray}}
\def\beg{\begin{align}}
\def\eeg{\end{align}}

\documentclass[final,5p,times,twocolumn,authoryear]{elsarticle}
\usepackage{amssymb}
\usepackage{lipsum}

\usepackage{leftindex}
\usepackage{mathtools, cuted}

\journal{Physics Letters B}

\begin{document}

\begin{frontmatter}

\title{Quantizing a Non-locally Massive 2-form Model}

\author[first]{Kumar Abhinav}
\affiliation[first]{organization={Centre for Theoretical Physics and Natural Philosophy, Nakhonsawan Studiorum for Advanced Studies},
            addressline={Mahidol University}, 
            city={Nakhonsawan},
            postcode={60130}, 
            country={Thailand}}

\begin{abstract}
A non-local yet gauge-invariantly massive 2-form model is considered that leads to local and unitary dynamics upon proper gauge-fixing. Since canonical momenta cannot be defined owing to the non-locality, consistent quantization of this system warrants particular care. We construct the Dirac brackets and covariant commutators for this model, followed by its path integral quantization, leading to consistent results. These treatments underline the most efficient way to extract the physical modes and the subsequent spectrum free from infrared ambiguity. Physically the massive 2-form represents a spin-1 excitation, with potential application to string theory and cosmology, that can mediate a screened interaction when topologically coupled to fermions.
\end{abstract}


\end{frontmatter}




\section{Introduction}
\label{sec:level1}
The interplay between gauge invariance and mass is a fundamental one. Though the absence of mass is taken to be synonymous with gauge invariance, it is well-known that mass and gauge invariance can coexist in the presence of conserved currents \citep{PhysRev.125.397,PhysRev.128.2425}. The reduction of $U(1)$ redundancy to a dynamical constraint upon introduction of mass \citep{refId0} manifests as extended gauge invariance over multiple fields \citep{Stueckelberg}. However, spontaneous symmetry breaking has been the most established mechanism for the gauge-invariant mass generation \citep{PhysRev.130.439,PhysRevLett.13.321,PhysRevLett.13.508,PhysRevLett.13.585}. In 2+1 dimensions, 1-form gauge mass can further be generated through self-interaction via the topological Chern-Simons term \citep{SCHONFELD1981157,PhysRevLett.48.975,DESER1982372}. The 3+1 dimensional counterpart of this topological mass generation \citep{CREMMER1974117,AURILIA1980265,Govindarajan_1982,doi:10.1142/S0217732391000580}, however, requires interaction between 1- and 2-forms as the $B\wedge F$ term \citep{PhysRevD.9.2273,PhysRev.187.1931}. 

Can a massive gauge theory exist in 3+1 dimensions with only one gauge field? A massive model with the Hodge dual $\leftindex^*F_{\mu\nu}$ of the 1-form field strength was proposed and duly quantized \citep{PhysRevD.1.2974}, that reduces to the Proca theory in the Lorentz gauge $\partial_\mu A^\mu=0$. Later, a gauge-invariant massive 1-form model was proposed \citep{PhysRevD.10.500},  
\be  
{\cal L}_A=-\frac{1}{4}F_{\mu\nu}\left(1+\frac{m^2}{\partial^2}\right)F^{\mu\nu},\quad F_{\mu\nu}=\partial_\mu A_\nu-\partial_\nu A_\mu,\label{10}
\ee
that was non-local.
This non-locality $\left(\partial^{-2}\right)$ does not hamper local and causal dynamics and leads to consistent quantization \citep{Vyas2016}. 

The present work proposes a similar non-local mass for the Abelian 2-form $B_{\mu\nu}=-B_{\nu\mu}$. The mass-less 2-form describes classical string dynamics \citep{PhysRevD.9.2273} and explains cosmic string interactions \citep{PhysRevD.35.1138} along with topological defects in superstrings \citep{ROHM1986454}. It further depicts Abelian superfluid vortices \citep{DAVIS1988219,PhysRevLett.63.2021,PhysRevD.40.4033} in type-II superconductivity \citep{doi:10.1142/S0217732390001062} and axionic charge allocation to black holes \citep{ALLEN199047}. Incorporating a gauge invariant mass would provide a finite interaction range to the 2-form. This should instigate novel effects, especially in cosmology, since the massless 2-form is responsible for inflatory physics \citep{Mavromatos2021a}, anisotropic inflation \citep{DebaprasadMaity_2004,Almeida_2019}, and primordial gravitational waves with dynamical dark energy \citep{Mavromatos2021}. 

The model we consider here describes a 2-form generalization of the non-local mass given as,
\bea
&&{\cal L}_B=\frac{1}{12}H_{\mu\nu\rho}\left(1+\frac{M^2}{\partial^2}\right)H^{\mu\nu\rho},\nonumber\\
&&H_{\mu\nu\rho}=\partial_\mu B_{\nu\rho}+\partial_\nu B_{\rho\mu}+\partial_\rho B_{\mu\nu}.\label{11}
\eea
It retains the original gauge-invariance $B_{\mu\nu}\to B_{\mu\nu}+\partial_\mu\Lambda_\nu-\partial_\nu\Lambda_\mu$ with a secondary redundancy: $\Lambda_\mu\to\Lambda_\mu+\partial_\mu\sigma$ and leads to local equations of motion under proper gauge-fixing. However, the non-locality does obstruct the construction of a canonical momentum, making the quantization of this model an intricate process. We achieve it here in multiple ways that require customized representations of this model tuned to the quantization process. The construction of canonical Dirac brackets requires a projective redefinition of the 2-form with a reduced $U(1)$ gauge structure. The complete gauge-invariance was utilized in constructing covariant quantum brackets that lead to a bounded spectrum in both covariant and non-covariant gauges. These brackets belong to a massive spin-1 3-vector following the Dirac treatment. Finally, a consistent path integral quantization required extending the model with a St\"uckelberg 1-form that led to decoupled ghosts with a proper BRST structure. As expected, this spin-1 field mediates a Yukawa-type interaction between suitably coupled fermions akin to the Meisner effect.

The manuscript is organized as follows. Section \ref{sec:level2} briefly reviews the massive 2-form formulation and the non-local 1-form model of \cite{Vyas2016}. The Classical degrees of freedom are obtained in Section \ref{sec:lavel3} highlighting the interplay of gauge-fixing and non-locality. Section \ref{sec:level4} contains the canonical Dirac analysis in a gauge-independent way, followed by the construction of quantum brackets in both covariant and non-covariant gauges, concluding with the path-integral treatment. We further compare these different approaches leading to the same physical outcomes. Conclusions and discussions regarding possible further aspects constitute Section \ref{sec:level5}.

\section{Massive gauge dynamics in 3+1 dimensions}
\label{sec:level2}
For the massless ($M=0$) 2-form theory, the equation of motion $\partial_\mu H^{\mu\nu\rho}=0$ enables the identification $H_{\mu\nu\rho}=\epsilon_{\mu\nu\rho\lambda}\partial^\lambda\varphi$ with a massless Klein Gordon scalar $\varphi$, possessing a single degree of freedom. By interacting with a 1-form the 2-form acquires a mass in the $B\wedge F$ model,
\be
{\cal L}_{\rm AB}=-\frac{1}{4}F_{\mu\nu}F^{\mu\nu}+\frac{1}{12}H_{\mu\nu\rho}H^{\mu\nu\rho}+\frac{g}{2}\epsilon^{\mu\nu\rho\sigma}B_{\mu\nu}F_{\rho\sigma}.\label{1}
\ee
The topological $B\wedge F$ term is metric-independent and does not contribute to the stress-energy tensor. As a result, the system obeys a massive Klein-Gordon dispersion describing a renormalizable spin-1 3-vector. Without any symmetry breaking, this mass generation is distinct from the Higgs mechanism \citep{doi:10.1142/S0217732391000580}. 

The $B\wedge F$ term appears via radiative correction due to massive fermions \citep{LEBLANC1994} or long-range spin currents \citep{Choudhury2015}. This is analogous to the radiatively induced Chern-Simons mass in 2+1 dimensions \citep{PhysRevLett.52.18,PhysRevD.29.2366,PhysRevLett.61.389}. The non-local 1-form model of Eq. \ref{10} effectively describes the St\"uckelberg theory by itself. It further appears through interaction with a confined pseudo-vector fermion current in 2+1 dimensions \citep{VYAS2017588} and in 1+1 dimensions as effective descriptions of both Schwinger model \citep{das2006field} and Ployakov's model of gravity \citep{doi:10.1142/S0217732387001130}. 

In the same way, the non-local 2-form Lagrangian ${\cal L}_{\rm B}$ effectively describes a tensorial St\"uckelberg generalization,
\bea
&&{\cal L}_{\rm BSt}=\frac{1}{12}H_{\mu\nu\rho}H^{\mu\nu\rho}-g^2B_{\mu\nu}B^{\mu\nu}+\frac{1}{2}B_{\mu\nu}\partial^{[\mu}D^{\nu]}\nonumber\\
&&\qquad\qquad\qquad\qquad\qquad-\frac{1}{16g^2}\partial_{[\mu}D_{\nu]}\partial^{[\mu}D^{\nu]},\label{2Stu1}
\eea
wherein the non-locality is compensated by the `St\"uckelberg 1-form' $\frac{1}{2g}D_\mu$. This massive 2-form also constitutes a dual description of type-II superconductivity through vortex-electron interaction \citep{MUKHERJEE2020168167} that leads to a $B\wedge F$ contribution at 1-loop. A 2+1-dimensional reduction of this dual model additionally leads to fractional statistics through flux attachment to spins \citep{MUKHERJEE2023116050} as an alternative to the Chern-Simons mechanism. The electrodynamics of vortices further leads to the non-local 2-form mass depicting Meisner screening \citep{Beekman2011}. Such occurrences of the massive 2-form make sense as the gauge-invariant mass can arise as an effective description \citep{Weinberg198051}. On the other hand, the well-behaved underlying theory suggests unitary and local quantum dynamics. The quantum versions of both massless 2-form \citep{Kaul} and $B\wedge F$ \citep{doi:10.1142/S021773239300369X} are known. In the latter case, the $B\wedge F$ coupling ensures that the gauge-dependent parts of the 1- and 2-form free propagators,
\bea
\Delta_{\mu\nu}&=&\frac{1}{\partial^2}\left[\eta_{\mu\nu}-(1-\xi)\frac{\partial_\mu\partial_\nu}{\partial^2}\right] \quad{\rm and} \nonumber\\
\Delta_{\mu\nu,\alpha\beta}&=&-\frac{1}{\partial^2}\left[\eta_{\mu[\alpha}\eta_{\beta]\nu}-(1-\zeta)\frac{\eta_{[\mu[\alpha}\partial_{\beta]}\partial_{\nu]}}{\partial^2}\right],
\label{11b}
\eea
do not contribute to the quantum theory. As a result, individual degrees of freedom of the free 1- and 2-form sectors simply combine into a massive 3-vector without any symmetry-breaking. As will be seen the non-locality in ${\cal L}_{\rm B}$, which disappears identically when the gauge is fixed, has a similar impact. subsequently, the physical dynamics can be extracted without requiring any auxiliary sector. Before proceeding further, below we briefly review the 1-form counterpart of this model.

\subsection{Quantized non-local 1-form theory}
The gauge-fixed version of the non-local 1-form model is \citep{Vyas2016},
\be
{\cal L}_A=-\frac{1}{4}F_{\mu\nu}\left(1+\frac{m^2}{\partial^2}\right)F^{\mu\nu}+B\partial_\mu A^\mu,\label{13}
\ee
where the Nakanishi-Lautrop auxiliary field $B$ implements the covariant gauge condition: $\partial\cdot A=0$. The Lagrange equations identify the auxiliary field as massless: $\partial^2B=0$, leading to the gauge dynamics,
\be
\left(\partial^2+m^2\right)\partial^2A_\mu=0.\label{15}
\ee
Since there are no more constraints, the 1-form has 3 independent components. However, there are two distinct modes in the theory. The massive mode $V_\mu=\partial^2A_\mu$ represents the physical spin-1 excitation responsible for a screened interaction that may support localized structures. The massless (Goldstone) mode $S_\mu=\left(\partial^2+4g^2\right)A_\mu$, on the other hand, conveniently decouples from the physical subspace. This is best seen in the covariant quantization formalism \citep{nakanishi1990covariant}, implemented through the unequal-time quantum bracket, 
\bea
&&[A_\mu(x),A_\nu(y)]=-i\eta_{\mu\nu}\Delta(x-y)+i\partial_\mu^x\partial_\nu^x\Delta'(x-y);\nonumber\\
&&\Delta(x)=-\frac{i}{(2\pi)^3}\int d^4p~{\rm sgn}(p_0)\delta(p^2-m^2){\rm e}^{-ip.x},\nonumber\\
&&\Delta'(x)=-\frac{1}{m^2}\Delta(x)+\frac{1}{m^2}\Delta(x, m=0).\label{17}
\eea
Herein the invariant commutator functions obey $\partial_{x,y}^2\Delta'(x-y)=\Delta(x-y)$. Subsequently, $V_\mu$ possesses a semi-positive-definite norm whereas $S_\mu$ corresponds to either zero or negative norm and annihilates the physical subspace.

The situation is much simpler in the temporal gauge $A_0=0$ as the equation of motion directly simplifies to that of a massive 3-vector: $\left(\partial^2+m^2\right)\boldsymbol{A}=0$. Similar physically meaningful dynamics can be expected for the non-local 2-form model.

\section{The Classical treatment}
\label{sec:lavel3}
The equation of motion that follows from the Lagrangian in Eq. \ref{11} has the form:
\be 
\left(1+\frac{4g^2}{\partial^2}\right)\partial_\mu H^{\mu\nu\rho}=0,\label{EOM}
\ee
with the re-parameterization $M=2g$ understood. The issue of non-locality in the Euler-Lagrange approach can be handled through auxiliary fields, like in Eq. \ref{2Stu1}, followed by on-shell substitution. 

The simplest way to identify the physical mode is in terms of the Hodge dual $\leftindex^*H_\mu$ of the field strength,
\be
\left(\partial^2+4g^2\right)\leftindex^*H_\mu=0,\quad\partial\cdot\leftindex^*H=0.\label{CA06}
\ee
depicting a massive 3-vector. Herein, the non-locality prevents identification with a scalar field, mirrored by the `St\"uckelberg 1-form' that contributes to the dynamics. In ${\cal L}_{\rm BSt}$ two independent gauge sectors exist,
\bea
&&B_{\mu\nu}\to B_{\mu\nu}+\partial_\mu\Lambda_\nu-\partial_\nu\Lambda_\mu,\quad D_\mu\to D_\mu+4g^2\Lambda_\mu,\nonumber\\
&&D_\mu\to D_\mu+\partial_\mu\lambda.\label{2Stu2}
\eea
They are consistent with both the equations of motion:
\bea 
&&\left(\partial^2+4g^2\right)B_{\mu\nu}-\partial_{[\mu}\left(\partial\cdot B\right)_{\nu]}=\partial_{[\mu}D_{\nu]},\nonumber\\
&&\partial^2D_\mu-\partial_\mu\partial\cdot D=4g^2\left(\partial\cdot B\right)_\mu,~\left(\partial\cdot B\right)_\mu=\partial^\nu B_{\nu\mu}. \label{2Stu4}
\eea 
These equations can be combined into that of a transverse gauge-invariant tensor $Z_{\mu\nu}=B_{\mu\nu}-\frac{1}{4g^2}\partial_{[\mu}D_{\nu]}$ obeying the massive Klein Gordon dispersion. Since the transverse-ness condition itself is further constrained as $\partial\cdot\left(\partial\cdot Z\right)=0$, there are 3 degrees of freedom left.

To carry out a canonical analysis, it is useful to manifest the gauge theory in a non-covariant form \citep{itzykson2012quantum}. Accordingly, the present model can be re-cast as,

\bea
&&{\cal L}_{\rm B}=\frac{1}{2}\left(\Pi_k\bigodot\Pi_k-\Phi\bigodot\Phi\right),\quad\bigodot=1+\frac{4g^2}{\partial^2},\nonumber\\
&&\boldsymbol{\Pi}=\partial_t\widetilde{\boldsymbol{B}}-\boldsymbol{\nabla}\times\overline{\boldsymbol{B}},\quad\Phi=\boldsymbol{\nabla} \cdot\widetilde{\boldsymbol{B}},\nonumber\\
&&\widetilde{B}_i=\frac{1}{2}\epsilon_{ijk}B_{jk},\quad\overline{B}_i=B_{0i},\quad i,j,k,=1,2,3.\label{CA01}
\eea
The equations of motion in terms of the field strengths take the form,
\be
\bigodot\left(\partial_t\boldsymbol{\Pi}-\boldsymbol{\nabla}\Phi\right)=0,\quad\bigodot\boldsymbol{\nabla}\times\boldsymbol{\Pi}=0,\label{CA03}
\ee
that yields massive dispersion for both fields. Then the 2-form Bianchi identity $\partial_t\Phi-\boldsymbol{\nabla}\cdot\boldsymbol{\Pi}=0$ again ensures a massive 3-vector. 

The gauge redundancy comes into effect in terms of the dynamical fields $\widetilde{\boldsymbol{B}}$ and $\overline{\boldsymbol{B}}$, the latter identified as a Lagrange multiplier. The corresponding equations of motion,
\bea
&&\partial_t^2\bigodot\widetilde{\boldsymbol{B}}=\boldsymbol{\nabla}\left(\boldsymbol{\nabla}\cdot\bigodot\widetilde{\boldsymbol{B}}\right)+\partial_t\boldsymbol{\nabla}\times\bigodot\overline{\boldsymbol{B}}\nonumber\\
{\rm and}~&&\partial_t\boldsymbol{\nabla}\times\bigodot\widetilde{\boldsymbol{B}}=\boldsymbol{\nabla}\times\boldsymbol{\nabla}\times\bigodot\overline{\boldsymbol{B}},\label{CA14}
\eea
are invariant under the gauge transformations,  
\be
\widetilde{\boldsymbol{B}}\rightarrow\widetilde{\boldsymbol{B}}+\boldsymbol{\nabla}\times\boldsymbol{\Lambda},~\overline{\boldsymbol{B}}\rightarrow\overline{\boldsymbol{B}}+\partial_t\boldsymbol{\Lambda}+\boldsymbol{\nabla}\Lambda,~\Lambda_\mu=\left(\Lambda,\,-\boldsymbol{\Lambda}\right).\label{CA15}
\ee
The standard covariant gauge condition $\left(\partial\cdot B\right)_\mu=0$ then translates to two mutually consistent conditions,
\be
\partial_t\overline{\boldsymbol{B}}+\boldsymbol{\nabla}\times\widetilde{\boldsymbol{B}}=0,\quad\boldsymbol{\nabla}\cdot\overline{\boldsymbol{B}}=0,\label{CA16}
\ee
which yields the massive dispersion $\left(\partial^2+4g^2\right)\mathfrak{B}=0$ where $\mathfrak{B}=\widetilde{\boldsymbol{B}},\,\overline{\boldsymbol{B}}$. Interdependence of the gauge conditions makes the transverse field $\overline{\boldsymbol{B}}$ vanish in the rest frame, leaving out $\widetilde{\boldsymbol{B}}$. In contrast, the transverse components of $\widetilde{\boldsymbol{B}}$ are determined by the redundant field $\overline{\boldsymbol{B}}$ in the massless case. 

The non-covariant gauge choice of $\overline{\boldsymbol{B}}=0$ expectantly \citep{itzykson2012quantum} simplifies the treatment as the first equation of motion directly leads to $\left(\partial^2+4g^2\right)\widetilde{\boldsymbol{B}}=0$. The second equation eludes to $\bigodot\boldsymbol{\nabla}\times\widetilde{\boldsymbol{B}}=0$, which does not restrict the dynamical field owing to the non-local operator. This `temporal' gauge is equivalent to the covariant one in the rest frame with $p^\mu=\left(2g,\,\boldsymbol{0}\right)$.

\section{Quantization of the 2-form model}
\label{sec:level4}
The off-shell non-locality makes the canonical description of the present system non-trivial. Meanwhile, the non-local term is required for a manifestly covariant treatment with a single gauge sector. Preferring this minimalist version of the model, we consider multiple quantization approaches requiring customized descriptions. We start with the Dirac approach, followed by covariant quantization in two distinct gauges. Finally, a standard path-integral treatment is carried out that requires auxiliary fields given the off-shell nature. With mutually consistent physics, they serve as templates for efficiently quantizing such non-local gauge models.

\subsection{Dirac Bracket analysis}
At a formal level, the non-locality in ${\cal L}_{\rm B}$ plagues the construction of conjugate momenta to the 2-form. Adopting the system of Eq. \ref{2Stu1} instead exchanges this problem with an additional auxiliary field ($D_\mu$). It is, however, more efficient to resolve the 2-form as $B_{\mu\nu}={\cal T}_{\mu\nu}+{\cal L}_{\mu\nu}$ in terms of projections,
\bea 
{\cal T}_{\mu\nu}=B_{\mu\nu}-{\cal L}_{\mu\nu},\quad{\cal L}_{\mu\nu}=\left(\eta_{\mu\alpha}\frac{\partial_\nu\partial_\beta}{\partial^2}+\eta_{\nu\beta}\frac{\partial_\mu\partial_\alpha}{\partial^2}\right)B^{\alpha\beta},\label{NN01}
\eea
which are also anti-symmetric. Here, ${\cal T}_{\mu\nu}$ is transverse $\partial^\mu{\cal T}_{\mu\nu}=0$ by construction, whereas ${\cal L}_{\mu\nu}=\partial_\mu\Lambda_\nu-\partial_\nu\Lambda_\mu$ is a pure gauge with $\Lambda^\mu=-\partial_\alpha B^{\alpha\mu}$. Thus $H_{\mu\nu\rho}$ is solely determined by gauge-independent component ${\cal T}_{\mu\nu}$. More importantly, the non-local term is now localized as $H_{\mu\nu\rho}\partial^{-2}H^{\mu\nu\rho}=-3{\cal T}_{\mu\nu}{\cal T}^{\mu\nu}$ and the Lagrangian becomes,
\be 
{\cal L}_{\rm B}=\frac{1}{12}H^{{\cal T}}_{\mu\nu\rho}H_{{\cal T}}^{\mu\nu\rho}-g^2{\cal T}_{\mu\nu}{\cal T}^{\mu\nu}.\label{NN04}
\ee 
Herein, the suffix/prefix indicates ${\cal T}_{\mu\nu}$ as the constituent field. 

In contrast to the usual Proca-type scenario, the massive field ${\cal T}_{\mu\nu}$ is transverse by construction, a condition that enters as an additional constraint in the Lagrangian,
\bea
&&{\cal L}'_{\rm B}=\frac{1}{12}H^{{\cal T}}_{\mu\nu\rho}H_{{\cal T}}^{\mu\nu\rho}-g^2{\cal T}_{\mu\nu}{\cal T}^{\mu\nu}+F^\nu\partial^\mu{\cal T}_{\mu\nu},\nonumber\\
&&\qquad=\frac{1}{2}\left[\left(\partial_t\widetilde{\boldsymbol{T}}-\nabla\times\overline{\boldsymbol{T}}\right)^2-\left(\nabla\cdot\widetilde{\boldsymbol{T}}\right)^2\right]-2g^2\left(\widetilde{\boldsymbol{T}}^2-\overline{\boldsymbol{T}}^2\right)\nonumber\\
&&\qquad\quad-\boldsymbol{F}\cdot\partial_t\overline{\boldsymbol{T}}-F_0\nabla\cdot\overline{\boldsymbol{T}}-\boldsymbol{F}\cdot\nabla\times\widetilde{\boldsymbol{T}}.\label{NN07}
\eea
As usual, the 2-form can further be resolved into 3-vectors $\widetilde{T}_i=(1/2)\epsilon_{ijk}{\cal T}_{jk}$ and $\overline{T}_i={\cal T}_{0i}$ with $F_\mu=\left(F_0,-\boldsymbol{F}\right)$ being the Lagrange multiplier. The conjugate momenta for the vector fields are,
\be 
\widetilde{\Pi}_i=\frac{\partial{\cal L}'_{\rm B}}{\partial\partial_t\widetilde{T}_i}=\partial_t\widetilde{T}_i-\left(\nabla\times\overline{\boldsymbol{T}}\right)_i,\quad\overline{\Pi}_i=\frac{\partial{\cal L}'_{\rm B}}{\partial\partial_t\overline{T}_i}=-F_i.\label{NN08}
\ee 
Here, the transverseness condition protects $\overline{\boldsymbol{T}}$ from being redundant with the Lagrange multiplier vector serving as its conjugate. Through the standard Legendre procedure, the canonical Hamiltonian is obtained as,
\bea 
&&{\cal H}_{\rm can}=\frac{1}{2}\widetilde{\boldsymbol{\Pi}}^2+\frac{1}{2}\left(\nabla\cdot\widetilde{\boldsymbol{T}}\right)^2+\widetilde{\boldsymbol{\Pi}}\cdot\nabla\times\overline{\boldsymbol{T}}\nonumber\\
&&\quad\qquad+2g^2\left(\widetilde{\boldsymbol{T}}^2-\overline{\boldsymbol{T}}^2\right)-\overline{\boldsymbol{\Pi}}\cdot\nabla\times\widetilde{\boldsymbol{T}}+F_0\nabla\cdot\overline{\boldsymbol{T}}.\label{NN09}
\eea
Since $F_0$ is the only redundant field, the primary Hamiltonian of this system,
\be 
{\cal H}_{\rm P}={\cal H}_{\rm can}+\lambda_0\Pi^F_0,\label{NN010}
\ee 
carries a single Lagrange multiplier $\lambda_0$.

The standard Dirac procedure with ${\cal H}_{\rm P}$ is straightforward, leading to the Dirac brackets,
\bea 
&&\left\{\widetilde{T}_i(\boldsymbol{x},t),\,\widetilde{\Pi}_j(\boldsymbol{y},t)\right\}_{\rm D}=\delta_{ij}\delta^3(\boldsymbol{x}-\boldsymbol{y}),\nonumber\\
&&\left\{\overline{T}_i(\boldsymbol{x},t),\,\overline{\Pi}_j(\boldsymbol{y},t)\right\}_{\rm D}=\left(\delta_{ij}-\frac{\partial_i\partial_j}{\boldsymbol{\nabla}^2}\right)\delta^3(\boldsymbol{x}-\boldsymbol{y}),\nonumber\\
&&\left\{F_0(\boldsymbol{x},t),\,\Pi^F_0(\boldsymbol{y},t)\right\}_{\rm D}=0,\label{NN019}
\eea
that are consistent with all the secondary constraints (see \ref{app:3}). The transverseness $\boldsymbol{\nabla}\cdot\overline{\boldsymbol{T}}=0$ follows from the condition $\partial^\mu{\cal T}_{\mu\nu}=0$. Its other implication $\partial_t\overline{\boldsymbol{T}}+\boldsymbol{\nabla}\times\widetilde{\boldsymbol{T}}=0$ connects the transverse components of the two modes, leaving out correct degrees of freedom.   

In this projective representation, only a residual $U(1)$ redundancy remains (\ref{app:3}). Upon imposing the secondary constraints, the primary Hamiltonian takes the form,
\bea
&&{\cal H}_{\rm P}=\frac{1}{2}\widetilde{\boldsymbol{\Pi}}^2+\frac{1}{2}\left(\nabla\cdot\widetilde{\boldsymbol{T}}\right)^2+\widetilde{\boldsymbol{\Pi}}\cdot\nabla\times\overline{\boldsymbol{T}}\nonumber\\
&&\qquad\qquad\quad-\overline{\boldsymbol{\Pi}}\cdot\nabla\times\widetilde{\boldsymbol{T}}+2g^2\left(\widetilde{\boldsymbol{T}}^2-\overline{\boldsymbol{T}}^2\right).\label{NN020}
\eea
Only the longitudinal projection of $\widetilde{\boldsymbol{T}}$ would contribute to the dynamics if the mass is absent. This component is gauge-independently connected to that of the original 2-form: $\boldsymbol{\nabla}\cdot\widetilde{\boldsymbol{T}}=\boldsymbol{\nabla}\cdot\widetilde{\boldsymbol{B}}$. 

The present Dirac approach with gauge-independent fields ${\cal T}_{\mu\nu}$  is a way to tackle non-locally massive gauge models without invoking auxiliary dynamics \citep{Su}. It may be conjectured that if gauge invariance prevails with the non-local mass, a projective representation exists, leading to a consistent canonical structure.

\subsection{Gauge-invariant quantization}
Since the on-shell dynamics are locally subjected to gauge fixing, covariant quantum brackets can directly be constructed for the 2-form $B_{\mu\nu}$. Here, we follow the construction based on on-shell dynamics \citep{nakanishi1990covariant}, demonstrating gauge-invariance through covariant and non-covariant gauge conditions.

\subsubsection{In covariant gauge}
We begin by implementing the covariant $R_\xi$ prescription to the Lagrangian as,
\be 
{\cal L}_{\rm B}=\frac{1}{12}H_{\mu\nu\rho}\left(1+\frac{4g^2}{\partial^2}\right)H^{\mu\nu\rho}-\frac{\xi}{2}\left(\partial_\mu B^{\mu\nu}\right)^2.
\ee
In this Abelian case, the compensating ghost sector decouples as will be seen later. now the equation of motion takes the form, 
\be 
\left(1+\frac{4g^2}{\partial^2}\right)\partial_\mu H^{\mu\alpha\beta}=\xi\partial^{[\alpha}\left(\partial\cdot B\right)^{\beta]}.\label{EOMxi01}
\ee
It is easy to get rid of the gauge parameter $\xi$ revealing both massive and massless modes,
\be
\partial^2\left(\partial^2+4g^2\right)B_{\mu\nu}=0,\label{EOMxi02}
\ee
subjected to the condition $\partial^2\left(\partial\cdot B\right)_\mu=0$. This condition doubles as the covariant gauge for the massive mode $\partial^2B_{\mu\nu}$. The $R_\xi$ gauge propagator for $B_{\mu\nu}$,
\be
K_{\mu\nu,\alpha\beta}=\frac{1}{\partial^2+4g^2}\left[\eta_{\mu[\alpha}\eta_{\beta]\nu}-(1+\xi)\frac{\eta_{[\mu[\alpha}\partial_{\beta]}\partial_{\nu]}}{\partial^2}\right],\label{29}
\ee 
has a massive pole and a transverse gauge-independent part. Consequently, the coupling with a conserved external current $J_{\alpha\beta}$ is gauge-independent: $J^{\alpha\beta}(y)\partial_x^\mu K_{\mu\nu,\alpha\beta}\delta(x-y)=0$. In 3+1-dimension, a valid choice is the topologically conserved pseudo-current tensor $J_{\mu\nu}=\epsilon_{\mu\nu\alpha\beta}\partial^\alpha j^\beta$ where $j_\mu$ can represent charged fermions \citep{Choudhury2015,doi:10.1142/S0217732319500676}. Given the current is further time-independent the interaction potential,
\bea
&&V(\boldsymbol{x}_1,\boldsymbol{x}_2)=\frac{1}{2}\int\,d^3x\,J^{\mu\nu}(\boldsymbol{x})K_{\mu\nu,\alpha\beta}J^{\alpha\beta}\nonumber\\
&&=2\int\,d^3x\left(j_0\frac{1}{-\nabla^2+4g^2}j_0-j_i\frac{1}{-\nabla^2+4g^2}j_i\right),
\label{Pot3}
\eea
then describes a Yukawa-like interaction between localized charges and currents, akin to the Meisner-like effect. This conforms to the earlier observation of dynamical Meisner screening for Abrikosov vortices \citep{Beekman2011}.

Following the procedure of Klein-Gordon Divisor \citep{PhysRevD.1.2974}, we postulate the manifestly covariant unequal-time quantum bracket,
\bea
\left[B_{\mu\nu}(x),B_{\alpha\beta}(y)\right]&=&i\eta_{\mu[\alpha}\eta_{\beta]\nu}\Delta(x-y)-\frac{i}{\xi}\left(1+\frac{4g^2}{\partial_x^2}+\xi\right)\nonumber\\
&&\quad\times\eta_{[\mu[\alpha}\partial^x_{\beta]}\partial^x_{\nu]}\Delta'(x-y),\label{32}
\eea
that generalizes the 1-form case of Eq.s \ref{17}. Confinement of the non-locality to the gauge-dependent part of this bracket is instrumental for the local nature of the physical mode in the following. This gauge-dependent bracket is consistent with the equations of motion in Eq. \ref{EOMxi02} and the condition $\partial^2\left(\partial\cdot B\right)_\mu=0$, but not with the individual massive and massless dispersion. Hence it overlooks the constraints in the system and the physical mode is not entirely isolated since the $R_\xi$ prescription explicitly breaks gauge symmetry. Equivalently, the degrees of freedom are over-counted as the gauge condition cannot be imposed on quantum operators \citep{itzykson2012quantum}, and a Gupta-Bleuler condition,
\be 
\partial_\mu B^{\mu\nu\,(+)}\vert{\rm Phys.}\rangle=0,\label{GBL}
\ee
is needed on the physical subspace. 

The respective massive and massless modes,
\be
V_{\mu\nu}(x)=\partial^2B_{\mu\nu}(x),\quad S_{\mu\nu}(x)=\left(\partial^2+4g^2\right)B_{\mu\nu}(x),\label{34}
\ee
individually satisfy the gauge condition $\partial_\mu V^{\mu\nu}=0=\partial_\mu S^{\mu\nu}$ and constitute the quantum brackets,
\bea
\left[V_{\mu\nu}(x),V_{\alpha\beta}(y)\right]&=&i4g^2\left(4g^2\eta_{\mu[\alpha}\eta_{\beta]\nu}+\eta_{[\mu[\alpha}\partial^x_{\beta]}\partial^x_{\nu]}\right)\nonumber\\
&&\qquad\times\Delta(x-y),\nonumber\\
\left[S_{\mu\nu}(x),S_{\alpha\beta}(y)\right]&=&-\frac{i}{\xi}4g^2\left(1+\frac{4g^2}{\partial_x^2}+\xi\right)\eta_{[\mu[\alpha}\partial^x_{\beta]}\partial^x_{\nu]}\nonumber\\
&&\qquad\times\Delta(x-y,\,g=0),\nonumber\\
\left[V_{\mu\nu}(x),S_{\alpha\beta}(y)\right]&=&0.\label{37}
\eea
The massive and massless modes conveniently dissociate, with the latter's spuriousness reflected by its gauge-dependent, non-local commutator. Indeed, the massless mode decouples from the physical subspace as Eq. \ref{EOMxi01} implies,
\be 
S^{\mu\nu}=\left(1+\frac{4g^2}{\partial^2}+\xi\right)\partial^{[\mu}\left(\partial\cdot B\right)^{\nu]}.\label{ER001}
\ee
Therefore the massive mode entirely captures the gauge-independent dynamics, which can equivalently be captured through the Hodge dual as, 
\be
\left[\leftindex^*H_\mu(x),\leftindex^*H_\nu(y)\right]=\frac{i}{2}\left(4g^2\eta_{\mu\nu}-\partial_\mu\partial_\nu\right)\Delta(x-y),\label{42}
\ee
with semi-positive norm consistent with $\partial\cdot\leftindex^*H=0$. 

The manifest covariance is traded with the equal time limit to capture the on-shell spectrum, by going to the momentum space as,
\bea
&&b_{\mu\nu}({\bf q})=\frac{i}{\sqrt{2(2\pi)^3E_q}}\int d^3x~e^{iq.x}\overleftrightarrow{\partial_t}B_{\mu\nu}(x),\nonumber\\
&&E_q^2={\bf q}^2+4g^2.\label{39}
\eea
Therein the fundamental quantum bracket has the form,
\be
\left[b_{\mu\nu}({\bf q}),b^\dagger_{\alpha\beta}({\bf q}')\right]=\eta_{\mu[\alpha}\eta_{\beta]\nu}\delta^3({\bf q}-{\bf q}').\label{MomComCov}
\ee
The $\xi$-dependence duly disappears in this on-shell description. In terms of the 3-vector components,
\bea
&&\left[\widetilde{b}_i({\bf q}),\widetilde{b}^\dagger_j({\bf q}')\right]=\delta_{ij}\delta^3\left({\bf q}-{\bf q}'\right)=-\left[\overline{b}_i({\bf q}),\overline{b}^\dagger_j({\bf q}')\right],\nonumber\\
&&\left[\widetilde{b}_i({\bf q}),\overline{b}^\dagger_j({\bf q}')\right]=0.\label{MomComCov1}
\eea
The field $\widetilde{b}_i({\bf q})$ represents physical states with positive norm whereas the redundant field $\overline{b}_i({\bf q})$ has negative norm states. The covariant gauge condition,
\bea
&&\overline{\boldsymbol{b}}_{\rm T}({\bf q})\vert{\rm Phys.}\rangle=\frac{1}{E_q}\boldsymbol{q}\times\widetilde{\boldsymbol{b}}_{\rm T}({\bf q})\vert{\rm Phys.}\rangle,\nonumber\\
&&\overline{b}_{\rm L}({\bf q})\vert{\rm Phys.}\rangle=0.\label{EGF}
\eea
comes to rescue from these negative norm states. Here, the suffixes ${\rm L,T}$ stand for longitudinal and transverse components respectively. Since a rest frame exists the unphysical component $\overline{\boldsymbol{b}}_{\rm T}({\bf q})$ gets eliminated leaving behind the massive 3-vector $\widetilde{\boldsymbol{b}}({\bf q})$. 

The Hamiltonian in the covariant gauge has a local form,
\bea
&&H^C=\frac{1}{2}\int d^3x\Bigg[\Dot{\widetilde{\boldsymbol{B}}}\cdot\Dot{\widetilde{\boldsymbol{B}}}+\partial_i\widetilde{\boldsymbol{B}}\cdot\partial_i\widetilde{\boldsymbol{B}}+4g^2\widetilde{\boldsymbol{B}}\cdot\widetilde{\boldsymbol{B}}\nonumber\\
&&\qquad\qquad-\Dot{\overline{\boldsymbol{B}}}\cdot\Dot{\overline{\boldsymbol{B}}}-\partial_i\overline{\boldsymbol{B}}\cdot\partial_i\overline{\boldsymbol{B}}-4g^2\overline{\boldsymbol{B}}\cdot\overline{\boldsymbol{B}}\Bigg].\label{TE1}
\eea
On introducing a polarization basis $\left\{\varepsilon_{\mu\nu}\left(\boldsymbol{q};\,\lambda,\lambda'\right)\right\}$ (See \ref{app:4}) the 2-form can be expanded as,
\be 
b_{\mu\nu}\left(\boldsymbol{q}\right)=\sum_{\lambda,\lambda'=0}^3\,\beta_{\lambda\lambda'}\left(\boldsymbol{q}\right)\varepsilon_{\mu\nu}\left(\boldsymbol{q};\,\lambda,\lambda'\right).
\ee
Then the Hamiltonian has a normal-ordered expectation value in the physical subspace as,
\bea
&&\langle H^C\rangle_{\rm Phys.}=\int d^3q\,E_q\sum_{\lambda=1}^3\langle{\rm Phys.}\vert\widetilde{\beta}^\dagger_\lambda\widetilde{\beta}_\lambda\vert{\rm Phys.}\rangle,\nonumber\\
&&\quad\widetilde{\beta}_\lambda=\frac{1}{2}\epsilon_{\lambda\alpha\beta}\beta_{\alpha\beta},~\lambda,\alpha,\beta=1,2,3,\label{TE2}
\eea
denoting massive spin-1 dynamics.

On resolving in terms of the massive and massless modes,
\bea
&&\left[v_{\mu\nu}({\bf q}),v^\dagger_{\alpha\beta}({\bf q}')\right]=4g^2\left(4g^2\eta_{\mu[\alpha}\eta_{\beta]\nu}-\eta_{[\mu[\alpha}q_{\beta]}q_{\nu]}\right)\nonumber\\
&&\qquad\qquad\qquad\qquad\times\delta^3({\bf q}-{\bf q}'),\nonumber\\
&&\left[s_{\mu\nu}({\bf q}),s^\dagger_{\alpha\beta}({\bf q}')\right]=4g^2\eta_{[\mu[\alpha}q_{\beta]}q_{\nu]}\delta^3({\bf q}-{\bf q}'),\nonumber\\
&&\left[v_{\mu\nu}({\bf q}),s^\dagger_{\alpha\beta}({\bf q}')\right]=0,\label{40}
\eea
the two sectors dissociate\footnote{The bracket for $s_{\mu\nu}({\bf q})$ require the Landau gauge $\xi\to\infty$ since the bracket for $S_{\mu\nu(x)}$ is non-local.}. Subsequently, the 3-vector components $\left(\widetilde{\boldsymbol{v}}({\bf q}),\overline{\boldsymbol{v}}({\bf q})\right)$ and $\left(\widetilde{\boldsymbol{s}}({\bf q}),\overline{\boldsymbol{s}}({\bf q})\right)$ also dissociate among themselves. Both $\widetilde{\boldsymbol{s}}({\bf q})$ and $\overline{\boldsymbol{s}}({\bf q})$ turn out to have semi-negative norms:
\bea
\left[\widetilde{s}_i({\bf q}),\widetilde{s}^\dagger_j({\bf q}')\right]&=&4g^2\left(q_iq_j-{\bf q}^2\delta_{ij}\right)\delta^3\left({\bf q}-{\bf q}'\right)\nonumber\\
&=&\left[\overline{s}_i({\bf q}),\overline{s}^\dagger_j({\bf q}')\right].\label{NGFC}
\eea
As for the massive sector, both $\widetilde{\boldsymbol{v}}({\bf q})$ and $\overline{\boldsymbol{v}}({\bf q})$ correspond to semi-positive norms:
\bea
&&\left[\widetilde{v}_i({\bf q}),\widetilde{v}^\dagger_j({\bf q}')\right]=4g^2\left[E_q^2\delta_{ij}-q_iq_j\right]\delta^3({\bf q}-{\bf q}')\nonumber\\
&&\left[\overline{v}_i({\bf q}),\overline{v}^\dagger_j({\bf q}')\right]=4g^2\left[{\bf q}^2\delta_{ij}-q_iq_j\right]\delta^3\left({\bf q}-{\bf q}'\right).\label{MFC}
\eea
as expected. The longitudinal component of $\overline{\boldsymbol{v}}({\bf q})$ has a vanishing norm whereas its transverse components are related to those of $\widetilde{\boldsymbol{v}}({\bf q})$ as per Eq. \ref{EGF}. In general, all the components of $\widetilde{\ast}({\bf q})$ can never be replaced by those of $\overline{\ast}({\bf q})$. The states corresponding to $v_{\mu\nu}({\bf q})$ belong to the physical Fock space with semi-positive norms. Though ${\cal T}_{\mu\nu}=B_{\mu\nu}$ in the covariant gauge and $\overline{\boldsymbol{T}}$ is equated to the spurious mode $\overline{\boldsymbol{B}}$, it is only non-locally connected to the physical mode: ${\cal T}_{\mu\nu}=\partial^{-2}V_{\mu\nu}$. As a result, the Dirac brackets of Eq.s \ref{NN019} do not exactly correspond to the quantum ones in Eq.s \ref{MFC}, yet they furnish the same details of the physical mode.

\subsubsection{In non-covariant gauge}
We saw that the non-covariant gauge $\overline{\boldsymbol{B}}=0$ directly yields a massive Klein-Gordon field $\widetilde{\boldsymbol{B}}$ subjected to a subsidiary yet non-restrictive condition $\bigodot\boldsymbol{\nabla}\times\widetilde{\boldsymbol{B}}=0$. As the physical sub-space is now defined as $\overline{\boldsymbol{B}}^{(+)}\vert{\rm Phys.}\rangle=0$, a consistent quantum bracket can simply be,
\be 
\left[\widetilde{B}_i({\bf x},t),\,\Dot{\widetilde{B}}_j({\bf y},t)\right]=i\delta_{ij}\delta^3({\bf x}-{\bf y}).\label{ETQB0}
\ee 
In the massless limit, the subsidiary condition becomes effective and the bracket needs to have a longitudinal projection. 
The above bracket also follows from the covariant one in the Landau gauge and thus reproduces the momentum-space brackets of Eq.s \ref{MomComCov1}\footnote{The non-equal-time commutators satisfy the micro-causality condition as $\Delta(x-y)$ (and thus $\Delta'(x-y)$) vanishes for space-like separations. The equal-time condition serves as a special case of that and thus represents a causally definite condition for comparing quantization in the two gauges.}. The same physics coming out of different gauges indicates that the non-locality has been handled adequately while quantizing. Also, both gauges lead to the same gauge-independent commutator,
\be 
\left[\Pi_i({\bf x},t),\,\Phi({\bf y},t)\right]=i\partial^x_i\delta^3({\bf x}-{\bf y}).
\ee

That the inherent covariance of the system is retained in the non-covariant gauge quantization, is assured through the Dirac-Schwinger covariance condition \citep{itzykson2012quantum},

\be
\left[\mathfrak{T}_{00}({\bf x},t),\,\mathfrak{T}_{00}({\bf y},t)\right]=i\left(\mathfrak{T}_{0i}({\bf x},t)+\mathfrak{T}_{0i}({\bf y},t)\right)\partial^x_i\delta^3({\bf x}-{\bf y}).
\ee
It links the Hamiltonian and momentum densities that follow from the energy-momentum tensor,
\be 
\mathfrak{T}_{\mu\nu}=\partial_\mu\widetilde{\boldsymbol{B}}\cdot\partial_\nu\widetilde{\boldsymbol{B}}-\frac{1}{2}\eta_{\mu\nu}\left(\partial_\rho\widetilde{\boldsymbol{B}}\cdot\partial^\rho\widetilde{\boldsymbol{B}}-4g^2\widetilde{\boldsymbol{B}}\cdot\widetilde{\boldsymbol{B}}\right).
\ee
in this gauge. Since the Hamiltonian $\mathfrak{T}_{00}$ in Eq. \ref{TE1} is now free from $\overline{\boldsymbol{B}}$, it is much simpler to obtain the distribution of energy in Eq. \ref{TE2}. 

The gauge-independence can further be demonstrated by considering the fields,
\be 
\boldsymbol{X}=\partial_t^2\widetilde{\boldsymbol{B}}-\boldsymbol{\nabla}\left(\boldsymbol{\nabla}\cdot\widetilde{\boldsymbol{B}}\right)\quad{\rm and}\quad\boldsymbol{Y}=\partial_t\boldsymbol{\nabla}\times\widetilde{\boldsymbol{B}}.
\ee
For $\overline{\boldsymbol{B}}=0$, they satisfy the momentum-space commutators,
\bea
&&\left[x_i({\bf q}),\,x^\dagger_j({\bf q}')\right]=E_q^2\left[E_q^2\delta_{ij}-q_iq_j\right]\delta^3({\bf q}-{\bf q}'),\nonumber\\
&&\left[y_i({\bf q}),\,y^\dagger_j({\bf q}')\right]=E_q^2\left[\boldsymbol{q}^2\delta_{ij}-q_iq_j\right]\delta^3({\bf q}-{\bf q}'),
\eea
while being mutually independent: $\left[x_i({\bf q}),\,y^\dagger_j({\bf q}')\right]=0$. These commutators match those for the massive mode (Eq.s \ref{MFC}) in the covariant gauge. This correspondence owes to the non-local mass since $\overline{\boldsymbol{B}}=0$ in the rest frame in both gauges. $\boldsymbol{X}$ and $\boldsymbol{Y}$ are further related to the fields that constitute the Dirac brackets as $\widetilde{\boldsymbol{T}}=\partial^{-2}\boldsymbol{X}$ and $\overline{\boldsymbol{T}}=-\partial^{-2}\boldsymbol{Y}$, but non-locally. In the massless case, as can be expected, they vanish identically.

\subsubsection{Symmetry-breaking and non-locality}
To ensure the locality of the theory, a Nakanishi-Lautrup auxiliary field $C_\mu$ introduced as
\be 
-\frac{\xi}{2}\left(\partial_\mu B^{\mu\nu}\right)^2\rightarrow-C_\nu\partial_\mu B^{\mu\nu}+\frac{1}{2\xi}C_\mu C^\mu.\nonumber
\ee
`removes' the gauge parameter $\xi$ from the 2-form sector. The auxiliary field imbibes the gauge condition $C_\mu=\xi\left(\partial\cdot B\right)_\mu$ and therefore, is both massless and excluded from the physical subspace. Using the fundamental bracket,
\be
\left[C_\mu(x),B_{\alpha\beta}(y)\right]=i\left(\partial^2+4g^2\right)\eta_{\mu[\alpha}\partial_{\beta]}\Delta'(x-y),
\ee
$C_\mu$ duly dissociates with the massive mode: $\left[C_\mu(x),V_{\alpha\beta}(y)\right]=0$. With the massless mode, however, one component survives the equal-time limit,
\be
\left[C_i(\boldsymbol{x},t),\overline{S}_j(\boldsymbol{y},t)\right]=i4g^2\delta_{ij}\delta^3(\boldsymbol{x}-\boldsymbol{y}).
\ee
Indeed, $C_i$ constitute the conserved N\"other charge $Q_S(t)=-\left(i/4g^2\right)\int d^3xC_i(x)$ responsible for the shift symmetry associated with the massless mode: $\delta\overline{S}_j(x)=-i\theta\left[Q_S(t),\overline{S}_j(x)\right]=i\theta$ where $\theta\in\mathbb{R}$. Since this Nambu-Goldstone mode dissociates from the physical subspace with semi-negative norm (Eq. \ref{NGFC}), the massive physical sector has a well-defined infra-red limit despite the non-locality \citep{PhysRevD.10.500}\footnote{To appreciate the importance of gauge invariance, let us consider the non-local massive scalar model,
\be  
{\cal L}_\varphi=\left(1+\frac{\mu^2}{\partial^2}\right)\partial_\mu\varphi\partial^\mu\varphi.
\ee
It yields a local equation of motion $\partial^2\left(\partial^2+\mu^2\right)\varphi=0$ with a shift symmetry for the massless mode $S=\left(\partial^2+\mu^2\right)\varphi$. An auxiliary vector $J_\mu=\mu\partial^{-2}\partial_\mu\varphi$, that removes the non-locality, constitutes the corresponding generator $Q=-(i/\mu)\int d^3x\partial^2J_\mu(x)$ under the fundamental bracket $\left[\varphi(x),\varphi(y)\right]=i\Delta(x-y)$. The massive mode $V=\partial^2\varphi$, with a positive-definite norm $\left[V(x),V(y)\right]=i\mu^4\Delta(x-y)$, decouples from the Goldstone mode $\left[V(x),S(y)\right]=0$, the latter having a vanishing norm $\left[S(x),S(y)\right]=0$. Without a Gupta-Bleuler condition, the physical space carries the massless mode and the theory has a bad infra-red behavior.}.

\subsection{Path integral treatment}
The path-integral quantization works at the level of generating functional, making it valid up to all orders \citep{BECCHI1974344}. 
It is achieved by utilizing the gauge-invariance by isolating the redundant degrees of freedom as ghosts. However, the inherent off-shell nature mandates the removal of the non-locality through auxiliary fields. Both these requirements are fulfilled by the St\"uckelberg representation of Eq. \ref{2Stu1}. The choice $f_\mu:=\left(\partial\cdot B\right)_\mu=0$ in the shared gauge makes the 2-form massive. In the Faddeev-Popov prescription, the corresponding ghost Lagrangian is,
\be 
{\cal L}^B_{\rm Ghost}=\frac{1}{2}\partial_{[\mu}\overline{c}_{\nu]}\partial^{[\mu}c^{\nu]},\label{PI10}
\ee 
in addition to the gauge-fixing term,
\be 
{\cal L}^B_{\rm GF}=-\frac{\xi}{2}\left(\partial_\mu B^{\mu\nu}\right)^2.\label{PI11}
\ee
Higher-form massless gauge theories usually have redundant primary ghosts leading to the proliferation of ghost fields \citep{HATA1981527}. As the non-local mass is gauge-invariant, the present 2-form model is no exception since the vector (anti-)ghosts themselves are 1-forms,
\be 
\overline{c}_\mu\to\overline{c}_\mu+\partial_\mu\overline{\kappa},\quad c_\mu\to c_\mu+\partial_\mu\kappa,\label{PI12}
\ee
mandating the additional {\it ghost-of-ghost} Lagrangian,
\be 
{\cal L}^{\overline{c},c}_{\rm Ghost}=\partial_\mu\overline{d}\partial^\mu d+\partial_\mu\overline{e}\partial^\mu e.\label{PI13}
\ee
The Grassmann fields $\left(\overline{d},d,\overline{e},e\right)$ obey a statistics opposite to that of $\left(\overline{c}_\mu,\,c_\mu\right)$ \citep{doi:10.1142/6938}, and thus represent {\it positive} degrees of freedom. The gauge-fixing of the ghost sector is achieved through,
\be 
{\cal L}^{\overline{c},c}_{\rm GF}=-\frac{\eta}{2}\left(\partial\cdot\overline{c}\right)\left(\partial\cdot c\right).\label{PI14}
\ee
Further, the gauge-fixing condition $f_\mu$ itself is divergence-less. Implementing this restriction invokes another ghost term,
\be 
{\cal L}^{f}_{\rm Ghost}=\frac{1}{2}\partial_\mu f\partial^\mu f,\label{PI15}
\ee
having a scalar ghost $f$. Finally, the gauge redundancy of the $D_\mu$ sector is handled by the ghost action,
\be 
{\cal L}^{D}_{\rm Ghost}=\partial_\mu\overline{g}\partial^\mu g,\label{PI16}
\ee
with the corresponding gauge-fixing part,
\be 
{\cal L}^{D}_{\rm GF}=-\frac{\varepsilon}{2}\left(\partial\cdot D\right)^2.\label{PI17}
\ee

All these parts now combine into the complete Lagrangian,
\bea 
&&{\cal L}_B^{PI}={\cal L}_B+{\cal L}^B_{\rm Ghost}+{\cal L}^B_{\rm GF}+{\cal L}^{\overline{c},c}_{\rm Ghost}+{\cal L}^{\overline{c},c}_{\rm GF}+{\cal L}^{f}_{\rm Ghost}\nonumber\\
&&\qquad\quad+{\cal L}^{D}_{\rm Ghost}+{\cal L}^{D}_{\rm GF}+B_{\mu\nu}J^{\mu\nu}++D_\mu j^\mu\nonumber\\
&&\qquad=\frac{1}{12}H_{\mu\nu\rho}H^{\mu\nu\rho}-g^2B_{\mu\nu}B^{\mu\nu}+\frac{1}{2}B_{\mu\nu}\partial^{[\mu}D^{\nu]}\nonumber\\
&&\qquad\quad-\frac{1}{16g^2}\partial_{[\mu}D_{\nu]}\partial^{[\mu}D^{\nu]}+\frac{1}{2}\partial_{[\mu}\overline{c}_{\nu]}\partial^{[\mu}c^{\nu]}\nonumber\\
&&\qquad\quad-\frac{\xi}{2}\left(\partial_\mu B^{\mu\nu}\right)^2+\partial_\mu\overline{d}\partial^\mu d+\partial_\mu\overline{e}\partial^\mu e\nonumber\\
&&\qquad\quad-\frac{\eta}{2}\left(\partial\cdot\overline{c}\right)\left(\partial\cdot c\right)+\frac{1}{2}\partial_\mu f\partial^\mu f+\partial_\mu\overline{g}\partial^\mu g\nonumber\\
&&\qquad\quad-\frac{\varepsilon}{2}\left(\partial\cdot D\right)^2+B_{\mu\nu}J^{\mu\nu}+D_\mu j^\mu,\label{PI18}
\eea
with proper sources $\left(J_{\mu\nu},\,j\right)$ introduced for generating $n$-point functions. Here, $D_\mu$ imbibes a physical mode, unlike the Nakanishi-Lautrup fields, as it represents the non-locality. Correspondingly, the generating function has the form, 
\bea
&&{\cal Z}\left[J_{\mu\nu},\,j_\mu\right]={\cal N}_0\int{\cal D}\Big[B_{\mu\nu},D_\mu,\overline{c}_\mu,c_\mu,\overline{d},\,d,\nonumber\\
&&\qquad\qquad\qquad\overline{e},\,e,\,f,\,\overline{g},\,g\Big]
\exp\left(i\int d^4x~{\cal L}_B^{PI}\right),\label{PI19}
\eea  
with overall normalization ${\cal N}_0$. 

The anti-symmetry leaves $B_{\mu\nu}$ with 6 degrees of freedom whereas $D_\mu$ has 4. The (anti-)ghosts $\left(\overline{c}_\mu,c_\mu,f,\,\overline{g},\,g\right)$ posses $-4-4-1-1-1=-11$ degrees of freedom in total, whereas that number for (anti-)ghost of (anti-)ghosts $\left(\overline{d},\,d,\,\overline{e},\,e\right)$ is $1+1+1+1=4$. Therefore the total degrees of freedom
of the system is $6+4-11+4=3$ as required. The pole structure of the 2-point function,
\be
K^{\mu\nu,\alpha\beta}(x,y)=\frac{-i}{{\cal Z}\left[J_{\mu\nu}\right]}\frac{\delta^2{\cal Z}\left[J_{\mu\nu}\right]}{\delta J_{\mu\nu}(x)\delta J_{\alpha\beta}(y)}\bigg\rvert_{J_{\mu\nu}=0},\label{47}
\ee 
reveals the massive excitation identical to that in Eq. \ref{29}.

Since there are two distinct gauge sectors $\left(\Lambda_\mu,\lambda\right)$, different gauge choices warrant different treatments though their equivalence is readily obtained. The alternate choice $\left(\partial\cdot B\right)_\mu+D_\mu=0$ might feel appealing since it leads to massive dynamics for both $B_{\mu\nu}$ and $D_\mu$ with positive energies. The corresponding physical subspace satisfies,
\be 
\left[\left(\partial\cdot B\right)_\mu+D_\mu\right]\vert{\rm Phys}\rangle=0.
\ee
However, since the path integral quantization is off-shell this gauge is equivalent to choosing $\left(\partial\cdot B\right)_\mu=0$. To appreciate that, let us consider the gauge-fixing,
\bea
&&\left(\partial\cdot B\right)_\mu+D_\mu\to\left(\partial\cdot B\right)_\mu+D_\mu\nonumber\\
&&\qquad+\partial^2\Lambda_\mu-\partial_\mu\partial\cdot\Lambda+4g^2\Lambda_\mu+\partial_\mu\lambda=0.\label{GFLl1}
\eea
On taking a divergence: $\partial\cdot D+4g^2\partial\cdot\Lambda+\partial^2\lambda=0$ the parameter $\lambda$ can be substituted to yield,
\bea
&&\left(\partial\cdot B\right)_\mu+D_\mu-\frac{1}{\partial^2}\partial_\mu\partial\cdot D\nonumber\\
&&\qquad+\left(\partial^2+4g^2\right)\left(\Lambda_\mu-\frac{1}{\partial^2}\partial_\mu\partial\cdot\Lambda\right)=0.\label{GFL}
\eea
Effectively, the $\Lambda_\mu$ sector possesses a transverse gauge-fixing function $f_\mu=\left(\partial\cdot B\right)_\mu+D_\mu-\partial^{-2}\partial_\mu\partial\cdot D$. It requires bosonic ghosts of the form in Eq. \ref{PI15} and will further implement a fermionic vector ghost action,
\be 
{\cal L}^B_{\rm Ghost}=-\frac{1}{2}\overline{c}_\mu\left(1+\frac{4g^2}{\partial^2}\right)\left(\partial^2\eta^{\mu\nu}-\partial^\mu\partial^\nu\right)c_\nu,\label{GFcmcm}
\ee
replacing Eq. \ref{PI10}. Interestingly, the non-local factor now appears in the ghost sector as $f_\mu=0$ leaves out the longitudinal component of $D_\mu$. The redundant vector ghost sector will again require two sets of ghost-of-ghosts. The vector gauge parameter now has the expression,
\be 
\Lambda_\mu=\frac{1}{\partial^2+4g^2}\left[\left(\partial\cdot B\right)_\mu+D_\mu-\frac{1}{\partial^2}\partial_\mu\partial\cdot D\right].\label{LamT}
\ee
It is local yet transverse: $\partial\cdot\Lambda=0$, making the auxiliary 1-form sector independent as,
\be
\partial\cdot D\to\partial\cdot D+\partial^2\lambda=0.\label{GFl}
\ee
Therefore the number of (anti-)ghost components remains the same as before, yielding the same physical mode.

The gauge symmetry is now replaced by an extensive nilpotent (anti-)BRST symmetry. We list here a set of BRST transformations that leave the action in Eq. \ref{PI18} invariant,
\bea 
&&\delta B_{\mu\nu}=\partial_\mu c_\nu-\partial_\nu c_\mu,\quad\delta D_\mu=4g^2c_\mu+\partial_\mu g,\nonumber\\
&&\delta\overline{c}_\mu=-\xi\left(\partial\cdot B\right)_\mu+\partial_\mu d,\quad\delta c_\mu=\partial_\mu e,\nonumber\\
&&\delta\overline{d}=-\frac{\eta}{2}\partial\cdot\overline{c},\quad\delta d=0,\nonumber\\
&&\delta\overline{e}=-\frac{\eta}{2}\partial\cdot c,\quad\delta e=0,\nonumber\\
&&\delta\overline{g}=-\varepsilon\partial\cdot D,\quad\delta g=0,\label{BRST2}
\eea
where $\delta^2=0$ due to nilpotency. A similar set of complementing anti-BRST transformations also exists. Invariance under these transformations ensures quantization up to all orders \citep{BECCHI1974344}. The physical sector is now free from negative-norm states since all the (anti-)ghosts decouple from the Abelian fields.

\section{\label{sec:level5}Summary and Conclusions}
In summary, quantizing the non-local massive 2-form model provides unique challenges. Since localization comes after gauge-fixing the conjugate momenta could be defined only in a projective representation with a residual $U(1)$ redundancy. Consequently, the Dirac brackets depict massive spin-1 dynamics. For multiple gauge conditions, consistent quantum brackets could be constructed which reproduced this spin-1 mode with subsequent spectrum. In the covariant gauge, a spurious Goldstone mode needed careful removal whereas the non-covariant gauge exclusively retained the massive physical mode. Finally, the off-shell path-integral treatment required the generalized St\"uckelberg version, leading to correct degrees of freedom with proper BRST symmetry. 

Such a non-locality should be a recurring feature of a single-field description of massive gauge dynamics, needing customized treatments for particular quantization schemes. This becomes particularly true if manifest gauge invariance is to be maintained during quantization. However, an on-shell treatment is still possible without the auxiliary dynamics as the problematic massless mode becomes unphysical. The remaining spin-1 excitation can mediate a Meisner-like screened interaction among properly coupled fermions.

In 3+1 dimensions the massless 2-form is equivalent to a free scalar whereas the massive Kalb-Ramond field is connected to the Proca theory \citep{doi:10.1142/S0217732396001922}. Being gauge-invariant the present model is linked to the St\"uckelberg system, with potential for interesting physics in string theory and cosmology. We hope to study these aspects further, especially when interacting fermions are present. 

The non-local 1-form model recently described early anisotropic deceleration and late time acceleration of the Universe \citep{HAGHANI2021100817}. A larger class of cosmological models utilizes generalizations to Proca theory \citep{Heisenberg_2014} and beyond \citep{PhysRevD.95.104001,HEISENBERG2016617}. Such models have a larger parameter range than scalar cosmologies. The present model naturally qualifies for similar applications as the corresponding Lagrangian effectively takes the form,
\be
{\cal L}_B\to\frac{1}{2}\left(\partial_\mu\varphi\right)^2-M\varphi\partial\cdot K
+\frac{1}{2}K_\mu\left(\partial^2+M^2\right)K^\mu,\label{NE01}
\ee
where $M^2=4g^2$. The on-shell elimination of the scalar $\varphi$ leads to Proca-like dynamics for the vector $K_\mu$,
\be 
\left(\partial^2+M^2\right)K^\mu=0,\quad \partial\cdot K=0.
\ee
distinct from the massive Kalb-Ramond field, the latter being a combination of axion and Proca fields \citep{capanelli2023cosmological}. The present model may also serve as a dark matter candidate, and thereby possibly the role of 2-form during inflation \citep{Mavromatos2021}. We plan to study these aspects soon.

\section*{Acknowledgements}
The present manuscript is dedicated in the memory of Professors Trilochan Pradhan and Roman W. Jackiw. The author deeply appreciates valuable input from Dr. Vivek M. Vyas during the inception of this work. This work enjoys the support of Mahidol University.

\appendix

\section{Construction of Dirac brackets}
\label{app:3}
In the primary Hamiltonian of Eq. \ref{NN010} the usual Poisson bracket for the constituent fields $\left\{\psi\right\}=\left(\widetilde{\boldsymbol{T}},\overline{\boldsymbol{T}},F_0\right)$ with their conjugate momenta $\Pi_\psi$ have the form,
\be
\left\{\psi(\boldsymbol{x},t),\,\Pi_\psi(\boldsymbol{y},t)\right\}_{\rm PB}=\delta^3(\boldsymbol{x}-\boldsymbol{y}).
\ee
Herein the only primary constraint $\varphi^1=\Pi^F_0\approx 0$ vanishes on the constraint hyper-surface. Weak vanishing of its time-evolution leads to further constraints as,
\bea
&&\partial_t\varphi^1=\left\{\varphi^1,\,{\cal H}_{\rm P}\right\}_{\rm PB}=-\nabla\cdot\overline{\boldsymbol{T}}:=\varphi^2\approx 0.\nonumber\\
&&\Rightarrow\partial_t\varphi^2=\left\{\varphi^2,\,{\cal H}_{\rm P}\right\}_{\rm PB}=-\boldsymbol{\nabla}\cdot\boldsymbol{\nabla}\times\widetilde{\boldsymbol{T}}=0,\label{NN012}
\eea 
and the series terminates. $\varphi^{1,2}$ are both first class and the Lagrange multiplier $\lambda_0$ stays arbitrary as the gauge condition $\partial^\mu{\cal T}_{\mu\nu}=0$ has interdependent components. Subsequently, $F_\mu$ has a $U(1)$ redundancy as $F^\nu\partial^\mu{\cal T}_{\mu\nu}=-\frac{1}{2}\left(\partial^\mu F^\nu-\partial^\nu F^\mu\right){\cal T}_{\mu\nu}$ modulo total derivative\footnote{This is equivalent to the secondary gauge sector $\Lambda_\mu\to\Lambda_\mu+\partial_\mu\lambda$.}. The constraints are made second class through two gauge conditions,
\be 
\rho^1:=F_0\approx 0,\quad \rho^2:=-\boldsymbol{\nabla}\cdot\overline{\boldsymbol{\Pi}}\approx 0.
\ee
Collecting the constraints as $\left\{c^A\right\}=\left(\varphi^1,\varphi^2,\rho^1,\rho^2\right)$ the matrix of their Poisson brackets ${\cal C}_{AB}=\left\{c^A,\,c^B\right\}_{\rm PB}$ has the inverse,
\be 
\left\{{\cal C}^{-1}_{AB}\right\}=\begin{pmatrix}
    0 & 0 & 1 & 0 \\
    0 & 0 & 0 & \boldsymbol{\nabla}^{-2}\\
   -1 & 0 & 0 & 0 \\
    0 & -\boldsymbol{\nabla}^{-2} & 0 & 0 
\end{pmatrix}\delta^3(\boldsymbol{x}-\boldsymbol{y}).\label{NN017}
\ee
ensuring that there are no more constraints. Then the following the prescription,
\bea
&&\left\{M(\boldsymbol{x}),\,N(\boldsymbol{y})\right\}_{\rm D}=\left\{M(\boldsymbol{x}),\,N(\boldsymbol{y})\right\}_{\rm PB}-\int d^3w d^3z\nonumber\\
&&\quad\times\left\{M(\boldsymbol{x}),\,c^A(\boldsymbol{w})\right\}_{\rm PB}{\cal C}^{-1}_{AB}\left\{c^B(\boldsymbol{z}),\,N(\boldsymbol{y})\right\}_{\rm PB},\label{NN018}
\eea
leads to the non-trivial Dirac brackets of Eq. \ref{NN019}.

\section{Polarization basis for 2-form}
\label{app:4}
Given the anti-symmetry of $b_{\mu\nu}$ the polarization tensors can be chosen as,
\be
\varepsilon_{\mu\nu}\left(\boldsymbol{q};\,\lambda,\lambda'\right)=\varepsilon_{[\mu}\left(\boldsymbol{q};\,\lambda\right)\varepsilon_{\nu]}\left(\boldsymbol{q};\,\lambda'\right)
\ee
making the operators $b_{\lambda,\lambda'}\left(\boldsymbol{q}\right)$ anti-symmetric as well. A natural choice for the set of basis vectors $\varepsilon_{\mu}\left(\boldsymbol{q};\,\lambda\right)$ is that for a Proca field \citep{greiner1996field}\footnote{The vectors $\varepsilon_{\mu}\left(\boldsymbol{q};\,0/3\right)$ are undefined in the massless limit as $q_\mu$ is not normalizable. Then the covariant gauge spares only the longitudinal component of $\widetilde{\boldsymbol{\beta}}$.},
\bea
&&\varepsilon_{\mu}\left(\boldsymbol{q};\,0\right)=\frac{q_\mu}{2g},\quad\varepsilon_{\mu}\left(\boldsymbol{q};\,n\right)=\left(0,-\boldsymbol{\varepsilon}\left(\boldsymbol{q};\,n\right)\right),\nonumber\\
&&\varepsilon_{\mu}\left(\boldsymbol{q};\,3\right)=\left(\frac{\vert\boldsymbol{q}\vert}{2g},-\frac{q_0}{2g}\hat{\boldsymbol{q}}\right),\quad n=1,2.\label{BV1}
\eea
This is a complete orthonormal basis,
\bea
&&\sum_{\lambda=0}^3\eta_{\lambda\lambda}\varepsilon_{\mu}\left(\boldsymbol{q};\,\lambda\right)\varepsilon_{\nu}\left(\boldsymbol{q};\,\lambda\right)=\eta_{\mu\nu},\nonumber\\
&&\varepsilon\left(\boldsymbol{q};\,\lambda\right)\cdot\varepsilon\left(\boldsymbol{q};\,\lambda'\right)=\eta_{\lambda,\lambda'},~~ q\cdot\varepsilon\left(\boldsymbol{q};\,\lambda\right)=2g\eta_{0\lambda},
\eea
about the momentum $q_\mu$, so that $b_{\lambda,\lambda'}\left(\boldsymbol{q}\right)$ satisfy the same commutators as $b_{\mu\nu}\left(\boldsymbol{q}\right)$. The longitudinal $\left(\varepsilon_{\mu}\left(\boldsymbol{q};\,3\right)\right)$ and temporal $\left(\varepsilon_{\mu}\left(\boldsymbol{q};\,0\right)\right)$ components exist due to the non-zero mass. Unlike the Proca theory, the absence of any on-shell constraint allows all four basis vectors. Now the covariant gauge condition translates to,
\be
\sum_{\lambda=1}^3\bar{\beta}_\lambda\left(\boldsymbol{q}\right)\varepsilon^\mu_\lambda\left(\boldsymbol{q};\,\lambda'\right)\vert{\rm Phys.}\rangle=0,\quad\bar{\beta}_\lambda=\beta_{0\lambda}.\label{EGC}
\ee
It implies $\bar{\beta}_\lambda\left(\boldsymbol{q}\right)\vert{\rm Phys.}\rangle=0$ on considering the explicit basis vectors.

\bibliographystyle{elsarticle-harv} 
\bibliography{Ref}

\end{document}